\documentclass{elsart3p}

 \usepackage{graphics}
 \usepackage{graphicx}

\usepackage{amssymb}

\begin{document}

\begin{frontmatter}



\title{Point-contact spectroscopy of the borocarbide superconductor YNi$_{2}$B$_{2}$C}


\author[a]{Yu. G. Naidyuk},
\author[a]{D. L. Bashlakov},
\author[a]{I. K. Yanson},
\author[b]{G. Fuchs},
\author[b]{G. Behr},
\author[b]{D. Souptel},
\author[b]{S.-L. Drechsler}

\address[a]{B. Verkin Institute for Low Temperature Physics and
Engineering, National Academy  of Sciences of Ukraine,  47 Lenin
Ave., 61103, Kharkiv, Ukraine}
\address[b]{Leibniz-Institut f\"{u}r Festk\"{o}rper- und
Werkstoffforschung Dresden e.V., Postfach 270116, D-01171 Dresden,
Germany}

\begin{abstract}
Point-contact (PC) spectroscopy measurements on YNi$_{2}$B$_{2}$C
single crystals in the normal and superconducting (SC) state
($T_{c}\simeq$15.4\,K) for the main crystallographic directions
are reported. The PC study reveals the electron-phonon interaction
(EPI) function with a dominant maximum around 12\,meV and a
further weak structure (kink or shallow broad maximum) at higher
energy at about 50\,meV.
Other phonon maxima at 20, 24 and 32\,meV specified in the phonon
DOS of YNi$_{2}$B$_{2}$C by neutron measurements [PRB,
\textbf{55}, 9058 (1997)] are not resolved in the PC spectra
pointing out to the main role of the low energy phonon modes in
EPI. Directional study of the SC gap results in
$\Delta_{\rm{[100]}}\approx$ 1.5\,meV for the $a$- direction and
$\Delta_{\rm{[001]}}\approx$ 2.4\,meV along the $c$-axis which may
point to anisotropic and/or multiband behavior. Noteworthy, the
critical temperature $T_{\rm c}$ in all cases corresponds to that
of bulk samples. The value 2$\Delta_{\rm [001]}$/$k_{\rm B}T_{\rm
c}\approx$ 3.6 is close to the BCS one of 3.52, and the
temperature dependence $\Delta(T)$ is BCS-like, while for the
$a$-direction $\Delta(T)$ deviates from mean-field BCS behavior
above $T_c/2$. The directional variation in $\Delta$ can be
attributed to the multiband nature of the SC state in
YNi$_{2}$B$_{2}$C predicted 10 years ago (PRL, \textbf{80}, 1730
(1998)).

\end{abstract}

\begin{keyword}
YNi$_2$B$_2$C \sep borocarbides\sep point-contact spectroscopy
\sep superconducting gap \sep electron-phonon interaction

\PACS 72.10.Di, 74.45.+c, 74.70Dd
\end{keyword}
\end{frontmatter}

By point-contact (PC) researches both the superconducting (SC) gap
and the PC electron-phonon interaction (EPI) function
$\alpha^2_{\rm PC}F(\omega)$ can be determined by measuring the
first and second derivatives of the $I(V)$ characteristic of PC's
\cite{Naid}. Thus the PC spectroscopy is a powerful method to
study both the EPI spectra and the SC gap behavior.

In the family of borocarbide superconductors YNi$_{2}$B$_{2}$C is
remarkable because of its relative high critical temperature
$T_{c}\simeq$ 15.4\,K among nonoxide ternary compounds. In
particular, the nature of the attractive interaction and the SC
order parameter remain challenging. So far several thermodynamic,
transport and spectroscopic measurements give a clear evidence for
a notable anisotropy of the SC gap in this compound \cite{Izawa}.
Concerning the EPI studies there is not much information available
except the PC data cited in \cite{Naid}.

\begin{figure}
\begin{center}
\includegraphics[width=8cm,angle=0]{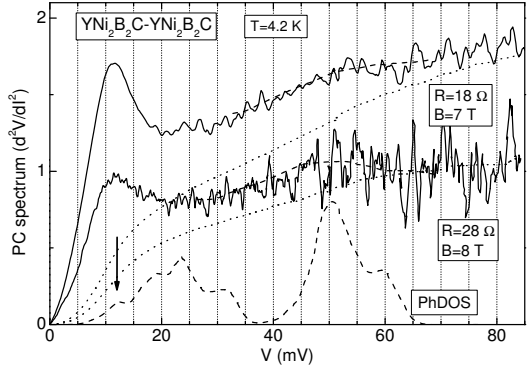}
\end{center}
\caption{(a) PC spectra of two YNi$_{2}$B$_{2}$C homocontacts
averaged for two polarities. The superconductivity is suppressed
by a magnetic field. Dotted curves show the tentative background
behavior. Dashed curves on the PC spectra are a guide for eyes to
improve the visualization of the maximum around 50\,mV. The bottom
curve shows the phonon DOS for YNi$_{2}$B$_{2}$C \cite{Gompf}.
} \label{f1}
\end{figure}

\begin{figure}
\includegraphics[width=8cm,angle=0]{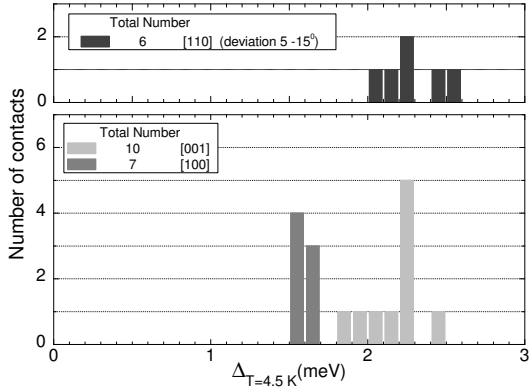}
\caption{Gap distribution for the three main directions in
YNi$_{2}$B$_{2}$C single crystal. } \label{f2}
\end{figure}

We have measured PC EPI spectra of YNi$_2$B$_2$C with a pronounced
phonon maxima at about 12\,mV and a broad maximum or shoulder
around 50\,mV (Fig.\,\ref{f1}). These peaks correspond to the
features seen in the phonon DOS \cite{Gompf}, however 20, 24 and
32\,mV phonon maxima are not resolved in the PC spectra. We have
obtained the PC EPI spectra for different directions and for
contacts with different SC gap $\Delta$, which is distributed (see
Fig.\,\ref{f2}) between 1.5\,meV and 2.5\,meV, however, no
qualitative difference between PC EPI spectra is observed. In all
cases, a more or less broad 12\,mV-maximum prevails in the PC EPI
spectra. This points out the main role of the low energy phonons
in EPI which contribution to the EPI constant $\lambda_{\rm
PC}=2\int\alpha^{2}_{\rm PC}F(\omega)\omega^{-1}d\omega$ is
estimated to about 90\%. PC EPI spectra were measured by
suppressing of superconductivity by magnetic field or temperature
to avoid features in the spectra due to SC gap. In this case we
did not found in the PC EPI spectra any "soft" modes at about
5\,mV mentioned, e.\,g., by Martinez-Samper et al. \cite{Izawa}.

\begin{figure}
\begin{center}
\includegraphics[width=7cm,angle=0]{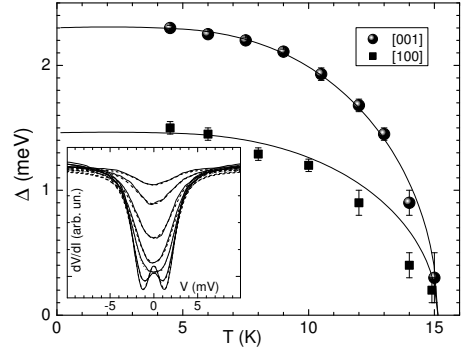}
\end{center}
\caption{Temperature dependence of the small and large gap in
YNi$_{2}$B$_{2}$C. Solid curves represent BCS-like behavior. Inset
shows example of $dV/dI$ curves (solid) for the small gap along
with fitting curves (dashed).}
 \label{f3}
\end{figure}

The SC gap $\Delta$ manifests itself in the $dV/dI$ curve of a
N-c-S contact as pronounced minima around $V\simeq\pm\Delta$ at
$T\ll T_{c}$. Such $dV/dI$ are presented in Fig.\,\ref{f3}(inset).
We have measured the gap distribution for the different
crystallographic directions in YNi$_{2}$B$_{2}$C shown in
Fig.\,\ref{f2}. The anisotropy in the distribution is clearly
seen: the small gap is characteristic for the a-axis, while along
the c-axis the gap is larger. Also the [110] direction has in
average the largest gap. Important is that for many of PCs with
different gap we have checked the critical temperature $T_{c, {\rm
PC}}$, which was always close to the bulk $T_{c}$. This avoids the
gap variation due to, i.\,e., the surface degradation.

The SC gap $\Delta$ and its temperature dependence are established
(Fig.\,\ref{f3}) from the fit of $dV/dI$. It is seen that
$\Delta(T)$ has BCS-type dependence, however the small gap
deviates from the BCS curve by approaching $T_{c}$. Similar
(small) gap behavior is characteristic for the multiband
superconductor MgB$_2$. For borocarbides, multiband
superconductivity has been firstly proposed already in 1998
\cite{Shulga}.


\begin{thebibliography}{00}

\bibitem{Naid}Yu. G.\ Naidyuk and I.K.\ Yanson, \textit{Point-Contact
Spectroscopy}, Springer Series in Solid-State Sciences, Vol.145
(Springer Science+Business Media, Inc, 2005).



\bibitem{Izawa}
K.\ Izawa et al.,
Phys.\ Rev. Lett. 89  (2002) 137006;
T.\ Park et al.,
Phys. Rev. Lett. 92 (2004) 237002;
P. Martinez-Samper et al.,
Phys. Rev. B 67 (2003) 014526;
P. Raychaudhuri et al.,
Phys. Rev. Lett. 93 (2004) 156802;
D. L. Bashlakov et al.,
Supercond. Sci. Technol. 18 (2005) 1094.

\bibitem{Gompf}F.\ Gompf et al.,
Phys. Rev. B 55 (1997) 9058.

\bibitem{Shulga} S.V. Shulga et al., Phys. Rev. Lett. 80 (1998)
1730.

\end{thebibliography}
\end{document}